\documentclass[runningheads]{llncs}
\usepackage{graphicx}
\usepackage{subfig}
\usepackage{cite}
\usepackage{enumitem}

\begin{document}
\title{Vehicle-To-Pedestrian Communication Feedback Module: A Study on Increasing Legibility, Public Acceptance and Trust\thanks{Supported by the Nevada NASA Space Grant Consortium, Grant No. 80NSSC20M00043.}}
\titlerunning{Vehicle-To-Pedestrian Communication Feedback Module}
\author{Melanie Schmidt-Wolf\inst{1}\orcidID{0000-0002-9056-1658} \and
David Feil-Seifer\inst{2}\orcidID{0000-0002-5502-7513}}

\authorrunning{M. Schmidt-Wolf and D. Feil-Seifer}

\institute{University of Nevada, Reno, 1664 N. Virginia Street, Reno, NV 89557-0171, USA\\
\email{mschmidtwolf@nevada.unr.edu, dave@cse.unr.edu}}

\maketitle              % typeset the header of the contribution
\begin{abstract}
Vehicle pedestrian communication is extremely important when developing autonomy for an autonomous vehicle. Enabling bidirectional nonverbal communication between pedestrians and autonomous vehicles will lead to an improvement of pedestrians' safety in autonomous driving. The autonomous vehicle should provide feedback to the human about what it is about to do. The user study presented in this paper investigated several possible options for an external vehicle display for effective nonverbal communication between an autonomous vehicle and a human. The result of this study will guide the development of the feedback module to optimize for public acceptance and trust in the autonomous vehicle's decision while being legible to the widest range of potential users.
The results of this study show that participants prefer symbols over text, lights and road projection. 
We plan to elaborate and focus on the selected interaction modes via Virtual Reality and in the real world in ongoing and future studies.

\end{abstract}

\section{Introduction}\label{intro}
Improving public acceptance, legibility, and trust in the autonomous vehicle's (AV's) decision is a significant open challenge for autonomous vehicles. Accidents are currently largely caused by human error \cite{wachenfeld2015freigabe}, which is why a major advantage of automated driving is the reduction and ideally the absence of human-induced accidents. Autonomous vehicles can eventually be expected to perform at high levels of precision without experiencing decreased performance like human drivers due to distraction or fatigue \cite{clamann2017evaluation}. Ultimately, these technologies will improve road safety, reduce injuries and save lives. 

However, interactions with high risk groups (i.e., pedestrians) remain a concern \cite{clamann2017evaluation}. The safety of all road users should be ensured to introduce autonomous driving in everyday life and substantially reduce traffic accidents. 
Vehicle pedestrian communication is extremely important when developing autonomy for that vehicle. The autonomous vehicle should provide feedback to the human about what it is about to do and what it would like the person to do. In this case the AV feedback is a replacement of the social signals of the human driver.
Figure 1 shows an example of a pedestrian indicating to cross the road, after which the AV provides corresponding feedback. 

In this paper, we present a study to identify which visual feedback module or combinations of feedback modules would increase most public acceptance, legibility, and trust in the autonomous vehicle's decision, and to identify preference. 

\begin{figure}[t]
\centerline{\includegraphics[width=0.5\textwidth]{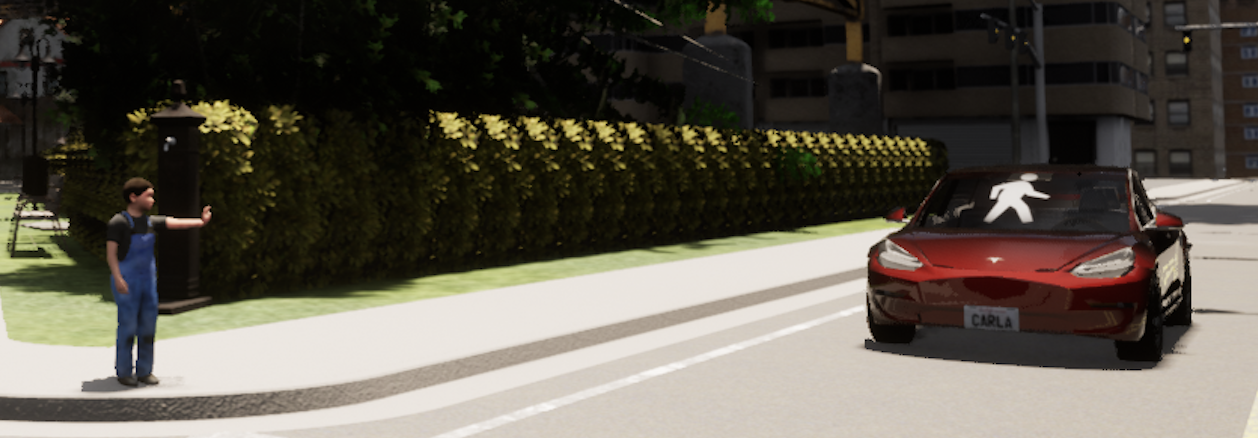}}
\caption{Example of vehicle human communication.}
\label{figureone}
\vspace{-1em}
\end{figure}
  
\section{Background} \label{Background}

The safety and efficiency of pedestrians crossing the road can be increased if AVs display their intention via an external human-machine interface (eHMI) to interact with pedestrians \cite{de2019external}.
Developers and researchers of autonomous vehicle technologies have proposed multiple types of displays, including digital road signs, text, audible chimes and voice instructions to communicate intent to pedestrians \cite{urmson2015pedestrian, de2019external, clamann2017evaluation, rover2018virtual}. 
In the following you can find a description of selected papers on studies of autonomous vehicle-pedestrian-communication-feedback modules. 
In De Clercq et al. \cite{de2019external}, different eHMI types were varied: baseline without eHMI, front brake lights, Knight Rider \cite{moody2013lone} animation (a light bar moves from left to right), smiley, a text which displays ``WALK". 
Lagström and Malmsten Lundgren \cite{lagstrom2016avip} developed a prototype HMI using a LED light strip in the top area on the windshield to communicate the vehicle’s current driving mode and intentions to the pedestrians. The vehicle communicated messages of either “automated driving mode,” “is about to yield,” “is resting,” or “is about to start,” which pedestrians understood after a short training \cite{lagstrom2016avip}. 
A study by Clamann et al. \cite{clamann2017evaluation} compared the effectiveness of various vehicle-to-pedestrian displays for street crossing. In this study, a prototype forward-facing display presenting information on a van investigated an advisory display with “Walk” and “Don’t Walk” symbols and an information display \cite{clamann2017evaluation}. 
Ackermann et al. considered in a video simulation study twenty HMI with projection, LED display and LED light strip, each with text-versus symbol-based message coding \cite{ackermann2019experimental}. 

Although there were several studies regarding a feedback module in the past, 
see also the reviews in \cite{rouchitsas2019external, carmona2021ehmi},
there is no clear indication about which feedback module would increase most the public acceptance and trust in the AV's decision.
In this paper, we expand on previous studies by comparing feedback module options via a questionnaire to identify which visual feedback module or combinations of feedback modules would increase most public acceptance, legibility, and trust in the autonomous vehicle's decision, and to identify preference.

\section{Method} \label{Method}
\subsection{Research Question}
The purpose of this study is to identify a feedback module to enable an autonomous vehicle to communicate with pedestrians, which increases most legibility, public acceptance and trust in the autonomous vehicle's decision, and to identify preference.
We selected the aspects public acceptance, legibility, and trust for the following reasons: 
\subsubsection{Legibility} The limited time pedestrians have to detect and interpret a signal is significant for the selection of a feedback module \cite{clamann2017evaluation}. Since the message displayed on an AV should be intuitive and concise, the message should be easy to understand \cite{rasouli2019autonomous}. 
\subsubsection{Public acceptance} The biggest obstacle in the mass adoption might not be technological, but public acceptance \cite{shariff2017psychological,  newcomb2012you}. Public acceptance is essential for the extensive adoption of AVs  \cite{yuen2020determinants}. 
\subsubsection{Trust} Trust has been identified as crucial to the successful design of AVs \cite{raats2020trusting}. The American Automobile Association (AAA) reports that only one in ten U.S. drivers would trust to ride in an AV, and 28\% of U.S. drivers are uncertain \cite{raats2020trusting, newsroom2020aaa}.

\subsection{Study Design}
This section describes the questionnaire used to identify a feedback module for communicating between a pedestrian and an autonomous vehicle, which increases most the legibility, public acceptance and trust in the autonomous vehicle's decision, and to identify preference. 
\subsubsection{Instruments}
To create the questionnaire we used Qualtrics XM, an online
survey tool. The participants were asked to watch different sections of videos and choose their
most likable option. These videos were designed using Blender (design elements), CARLA
(simulate AV environment), and Unreal Engine (bind physics to the elements). CARLA \cite{Dosovitskiy17} is based on Unreal Engine. In Unreal \mbox{Engine 4}, it is possible to create and modify objects, such as vehicles and the feedback displays. 
\subsubsection{Visualizations} 
\begin{figure*}[!b]
  \vspace{-2em}
    \centering
    \captionsetup[subfigure]{justification=centering}
        \subfloat[][Text: Walk]{\includegraphics[width=0.25\textwidth]{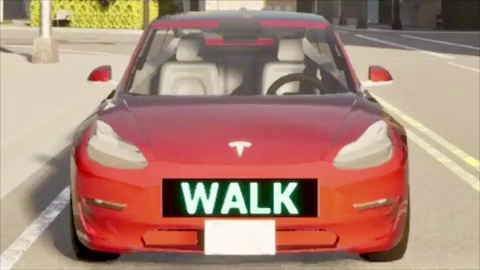}\label{text_walk}}
        \subfloat[][Text: Don't Walk]{\includegraphics[width=0.25\textwidth]{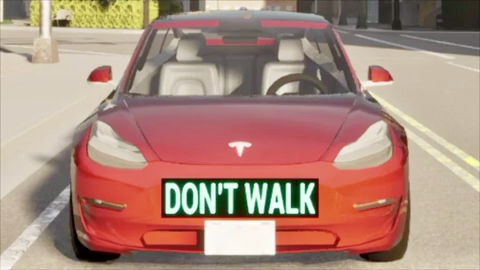}\label{text_dontwalk}}  
        \subfloat[][Symbol: Cross Advisory]{\includegraphics[width=0.25\textwidth]{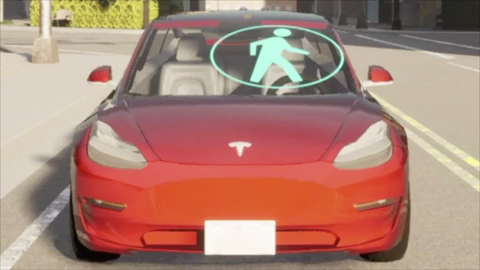}\label{symbol_crossadv}}  
        \subfloat[][Symbol: Stop Sign]{\includegraphics[width=0.25\textwidth]{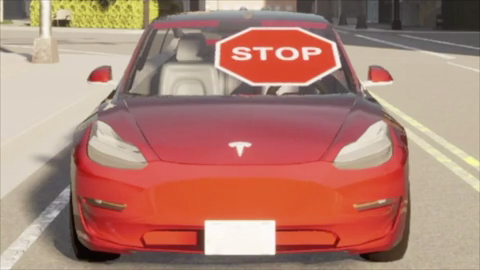}\label{symbol_stopsign}}
        \hfill
        \vspace{-1em}
        \subfloat[][Light: Green Front Brake Lights]{\includegraphics[width=0.25\textwidth]{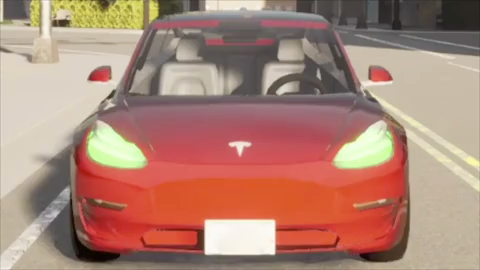}\label{light_green}}
        \subfloat[][Light: Static LED]{\includegraphics[width=0.25\textwidth]{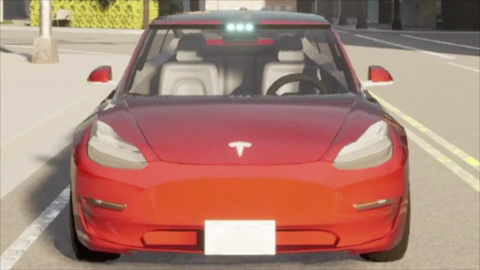}\label{light_staticled}}        
        \subfloat[][Road Projection: Zebra Crossing]{\includegraphics[width=0.25\textwidth]{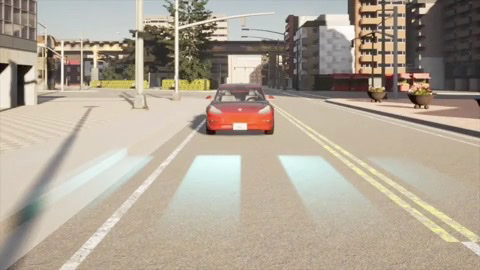}\label{proj_zebra}} \subfloat[][Road Projection: Lines far apart]{\includegraphics[width=0.25\textwidth]{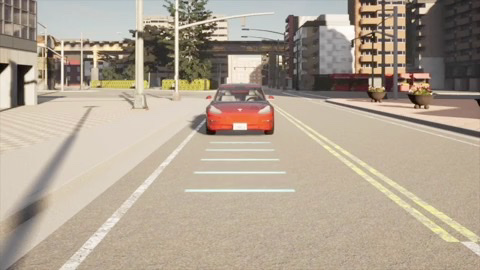}\label{proj_linesfar}} 
    \caption{Examples of visualizations of the four selected interaction modes: \protect\subref{text_walk} and \protect\subref{text_dontwalk} Text interaction mode  \protect\subref{symbol_crossadv} and \protect\subref{symbol_stopsign} Symbol interaction mode  \protect\subref{light_green} and
    \protect\subref{light_staticled} Light interaction mode  \protect\subref{proj_zebra} and \protect\subref{proj_linesfar} Road projection interaction mode with \protect\subref{text_walk}, \protect\subref{symbol_crossadv}, \protect\subref{light_green} and \protect\subref{proj_zebra} AV stops for the pedestrian \protect\subref{text_dontwalk}, \protect\subref{symbol_stopsign}, \protect\subref{light_staticled} and \protect\subref{proj_linesfar} AV does not stop for the pedestrian. In the questionnaire we ask about general concepts and add the videos/images for illustration purposes only. Perspective: The pedestrian's viewpoint is situated in front of the AV.}    
    \label{fig:visualization}
  \end{figure*}

We designed four sections to visualize the feedback module: text, symbols, lights and projections.
In Figure \ref{fig:visualization}, examples of interaction mode visualizations are displayed. 
For the perspective of the illustrations we chose to situate the pedestrian view in the front of the autonomous vehicle since it is most likely that pedestrians cross the road in front of the vehicle. For simplicity, and to not confuse participants with several perspectives, we selected one perspective to display the different interaction modes to the participants. 
The ideas for the different text, symbol, light and road projection interaction mode options result from literature review and brainstorming. 
The different concepts are acquired from prior research in this area.

\subsubsection{Questionnaire} 
Since there are a lot of different aspects to consider for a feedback module on an autonomous vehicle, we focused on making the questionnaire as simple and short as possible to answer our research question. 
We asked about the general concepts and added the videos and images for illustration purposes only. 
To decrease the influence of color we decided to use the color cyan in the illustrations as much as possible. Cyan or turquoise have been used in several studies regarding an AV-pedestrian-display, e.g., in \cite{de2019external, bazilinskyy2019survey}, due to being a neutral color in traffic and having good visibility. 

The questionnaire consisted of demographic related questions, and questions related to the AV’s visual feedback module and combinations of feedback modules. Based on the answers from the questionnaire, we analyzed the most favorable, publicly accepted,
legible and trusted visual feedback modules equipped by the AV. 
To explain the context of the questionnaire, participants were presented the following explanation: ``Imagine you are a pedestrian and you want to cross the road. An autonomous vehicle is approaching. You want to be sure that you can safely cross the road. But how will the autonomous vehicle tell you that you can cross? For this reason a visual feedback module will be used." 
The first part of the questionnaire is partially based on Schaefer's “Trust Perception Scale-HRI” \cite{schaefer2016measuring} with a 5-point Likert scale:
\begin{itemize}[noitemsep,topsep=0pt]
\item I believe the \_\_\_ interaction mode protects people from potential risks in the environment / looks friendly to the pedestrian / communicates clearly 
\item I prefer the \_\_\_ interaction mode over human-driver interaction. 
\end{itemize}

Those four questions were asked for each interaction mode Text, Symbols, Lights and Road Projections separately. To reduce bias we added randomization to the order of displaying the interaction modes as well as the order of the specific interaction mode options. 
To get a clear answer regarding our research question, participants ranked the interaction modes Text, Symbols, Lights and Road Projections regarding preference, legibility, public acceptance and trust directly: 
\begin{itemize}[noitemsep,topsep=0pt]
\item Please rank the interaction modes from the most legible interaction mode (1) to the least legible interaction mode (4) / from the interaction mode you trust the most (1) to the interaction mode you trust the least (4) / from the interaction mode you accept the most (1) to the interaction mode you accept the least (4) / in order of preference from your most preferred (1) to your least preferred (4) 
\end{itemize}

The demographic questions included age, gender, ethnicity and current level of education.

\subsection{Participants}
63 participants were recruited via flyers and social media to fill out the questionnaire online. The questionnaire has a duration of about 20 to 30 minutes. Twenty-six participants identified as female, 26 participants identified as male and the remaining 11 participants did not specify. Further, 13 participants identified as US-American and 29 participants identified as other nationality.

\section{Results and Discussion} \label{Results}
In this section we present the results followed by a discussion to  identify a feedback module to enable an autonomous vehicle to communicate with pedestrians, which increases most legibility, public acceptance and trust in the autonomous vehicle's decision, and to identify preference.

Since the Likert questions are ordinal we tested for normality with the Shapiro-Wilk normality test. Since the result of the Shapiro-Wilk normality test achieved a p-value that is less than $p < 0.05$ we cannot assume normality. Due to this result we used non-parametric tests and show the results of the questions with frequencies /percentages. 

\subsection{Legibility, Public Acceptance, Trust, and Preference}

\begin{figure}[!b]
\vspace{-1em}
    \centering
    \includegraphics[width=0.6\textwidth]{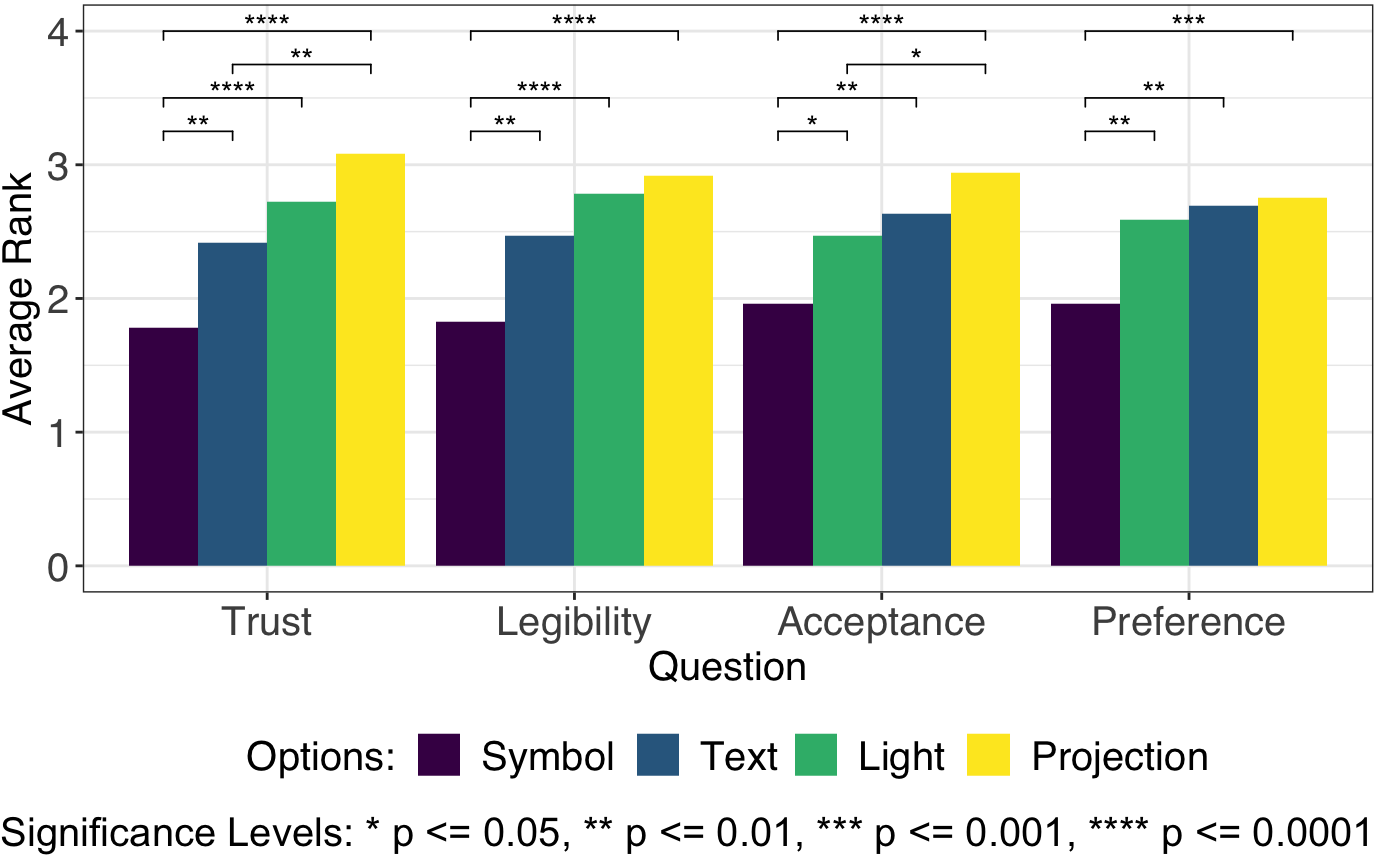}
    \caption{The result of ranking questions regarding trust, legibility, acceptance and preference shows that symbols should be selected as interaction mode due to significant differences between symbols and the interaction mode options text, light and road projection (lower is better).}
    \label{fig:ranking}
\end{figure}

To identify the feedback module which most increases legibility, public acceptance, and trust in the autonomous vehicle, and to identify the preferred feedback module,  we analyzed the ranking questions. 
The question ``Please rank the interaction modes in order of preference from your most preferred (1) to your least preferred (4)" resulted in the following average rank order: Symbol, Light, Text, Projection.
The question ``Please rank the interaction modes from the most legible interaction mode (1) to the least legible interaction mode (4)" resulted in the following average rank order: Symbol, Text, Light, Projection.
The question ``Please rank the interaction modes from the interaction mode you trust the most (1) to the interaction mode you trust the least (4)" resulted in the following average rank order: Symbol, Text, Light, Projection.
The question ``Please rank the interaction modes from the interaction mode you accept the most (1) to the interaction mode you accept the least (4)" resulted in the following average rank order: Symbol, Light, Text, Projection. 
Further, we tested each question for significance with the Kruskal-Wallis test, which showed that there are significant differences between the options. 

Figure \ref{fig:ranking} shows a visualization of the ranking questions results and the results of the pairwise comparisons with the Mann–Whitney U test. The Mann–Whitney U test shows a significant difference between symbols and the interaction mode options text, light and road projection with symbols ranked the highest.
\begin{figure*}[!b]
    \vspace{-1em}
    \captionsetup[subfigure]{justification=centering}
    \centering
    \subfloat{\includegraphics[width=0.85\textwidth]{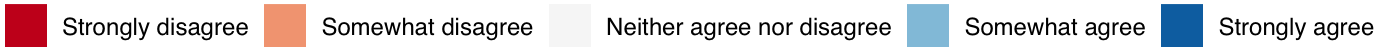}}  
    \setcounter{subfigure}{0}
    \begin{minipage}{.5\linewidth}
    \centering
    \subfloat[][I believe the \_\_\_ interaction mode protects people from potential risks in the environment]{\includegraphics[width=1.0\textwidth]{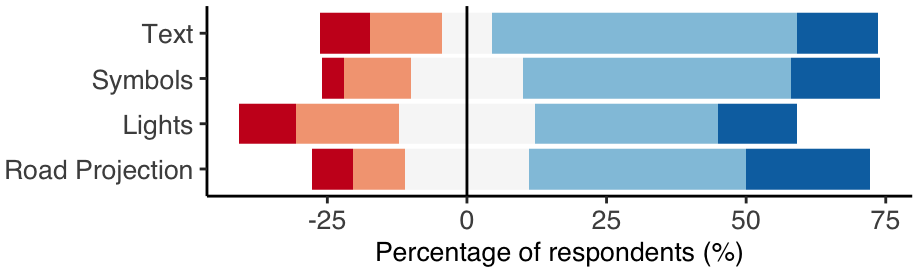}\label{likert_protects}}
    \end{minipage}%
    \begin{minipage}{.5\linewidth}
    \centering
    \vspace{-1em}
    \subfloat[][I believe the \_\_\_ interaction mode looks friendly to the pedestrian]{\includegraphics[width=1.0\textwidth]{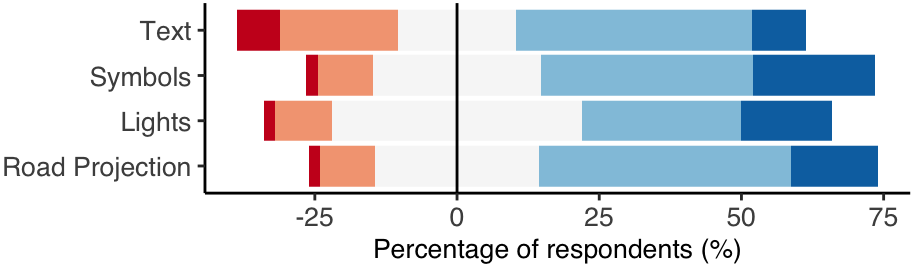}\label{likert_friendly}}
    \end{minipage}\par\medskip
    \centering
    \begin{minipage}{.5\linewidth}
    \centering
    \subfloat[][I believe the \_\_\_ interaction mode communicates clearly]{\includegraphics[width=1.0\textwidth]{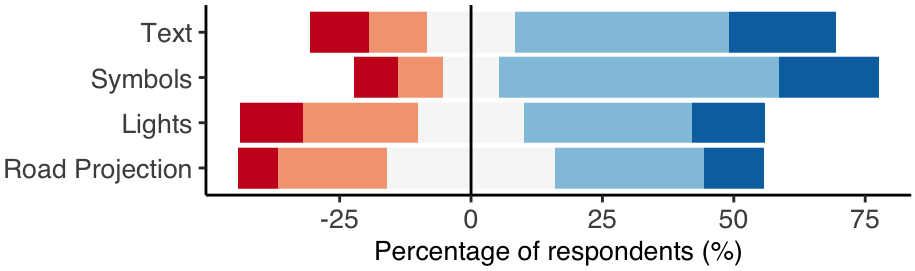}\label{likert_communicates}}
    \end{minipage}%
    \begin{minipage}{.5\linewidth}
    \centering
    \subfloat[][I prefer the \_\_\_ interaction mode over human-driver interaction]{\includegraphics[width=1.0\textwidth]{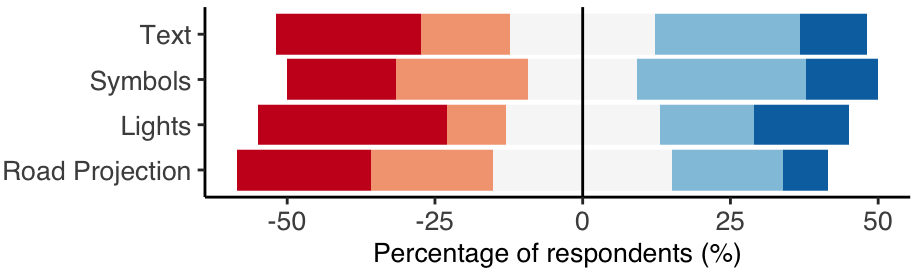}\label{likert_driver}} 
    \end{minipage}\par\medskip
    \centering  
    \vspace{-1em}
    \caption{
    Assessment by participants of the questions \protect\subref{likert_protects}, \protect\subref{likert_friendly}, \protect\subref{likert_communicates} and \protect\subref{likert_driver}
    with a 5-point Likert scale for the four selected interaction modes text, symbols, lights and road projection with resulting ranking in descending order: \protect\subref{likert_protects}  Text, symbols, road projection, lights
    \protect\subref{likert_friendly} Road projection, symbols, text, lights
    \protect\subref{likert_communicates} Symbols, text, lights, road projection
    \protect\subref{likert_driver} Symbols, text, lights, road projection. Participants preferred symbols, followed by text, lights and road projection.
    }
    \label{fig:likert}
\end{figure*}
Further, we analyzed the Likert questions sorted by interaction modes, see Figure \ref{fig:likert}. Regarding the question ``I believe the \_\_\_ interaction mode protects people from potential risks in the environment" participants agreed with the text interaction mode the most, followed by symbols, road projection and lights. Regarding this question the Kruskal-Wallis test showed no significance. ``I believe the \_\_\_ interaction mode looks friendly to the pedestrian" led to most agreement for the road projection interaction mode, followed by symbols, text and lights. The Kruskal-Wallis test showed no significance. The question if the participant believes that the interaction mode communicates clearly led to most agreement for the symbols interaction mode, followed by text, lights and road projection. Here, the Kruskal-Wallis test showed significance. Further, the Mann–Whitney U test resulted in pairwise significant differences between symbols and road projection, and symbols and lights. Furthermore, the question if the participant prefers the interaction mode over human-driver interaction led to most agreement for the symbol interaction mode, followed by text, light and road projection. The Kruskal-Wallis test showed no significance. 
Taking all results together regarding which feedback module increases most legibility, public acceptance and trust in the autonomous vehicle’s decision, and to identify preference, participants selected symbols, followed by text, lights and road projections.

\subsection{Specific interaction mode options} 
  
\begin{figure*}[!h]
    \captionsetup[subfigure]{justification=centering}
    \centering
    \begin{minipage}{1.0\linewidth}
    \centering
    \subfloat{\includegraphics[width=0.85\textwidth]{Likert/LegendLikert.png}} 
    \end{minipage}\par\medskip
    \vspace{-0.5em}
    \begin{minipage}{1.0\linewidth}
    \centering
    \subfloat{\includegraphics[width=0.35\textwidth]{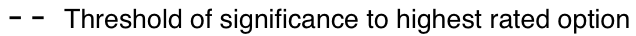}} 
    \end{minipage}\par\medskip
    \setcounter{subfigure}{0}
    \begin{minipage}{.5\linewidth}
    \centering
    \subfloat[][I believe the \_\_\_ interaction mode protects people from potential risks in the environment]{\includegraphics[width=1.0\textwidth]{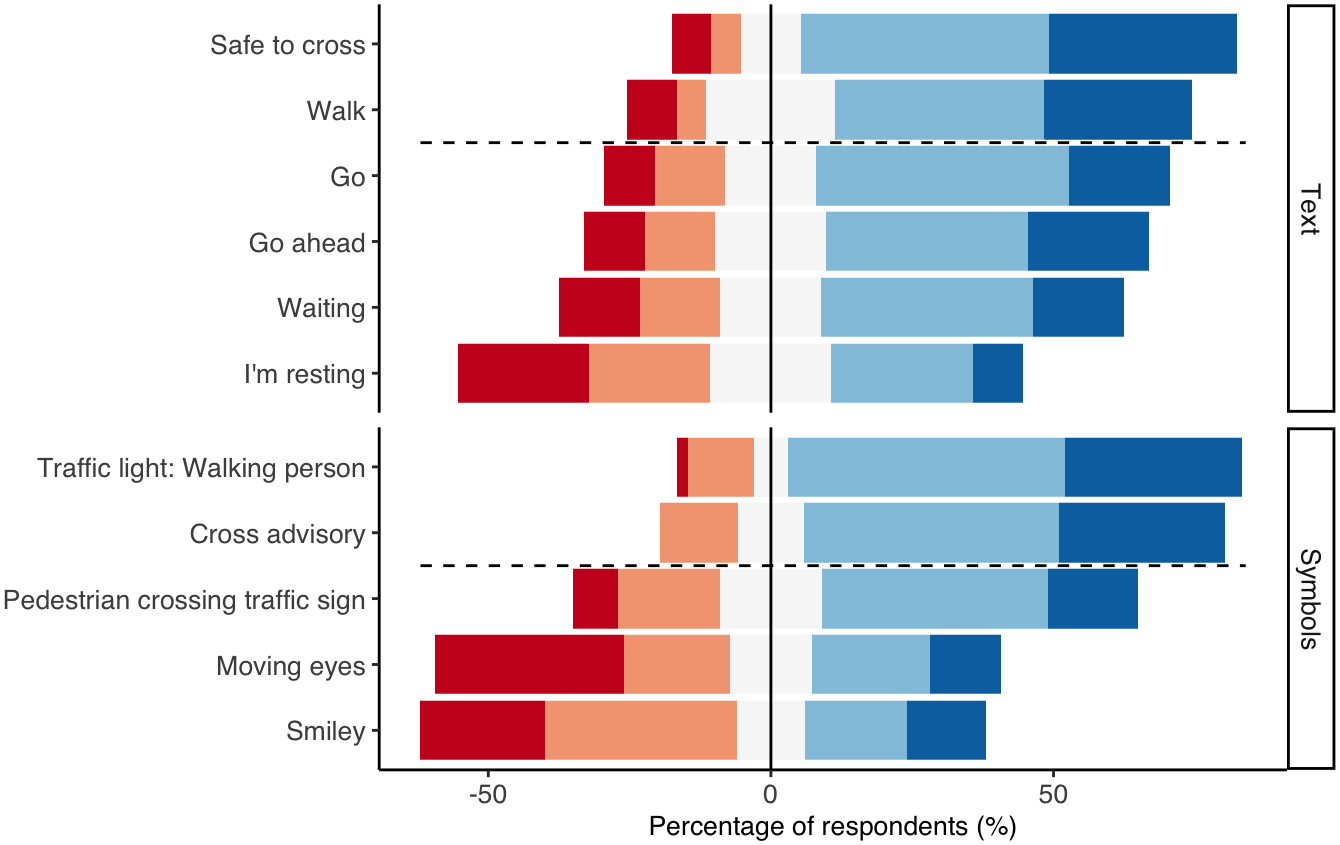}\label{likert_ts_protects}}
    \end{minipage}%
    \begin{minipage}{.5\linewidth}
    \centering
    \vspace{-1em}
    \subfloat[][I believe the \_\_\_ interaction mode looks friendly to the pedestrian]{\includegraphics[width=1.0\textwidth]{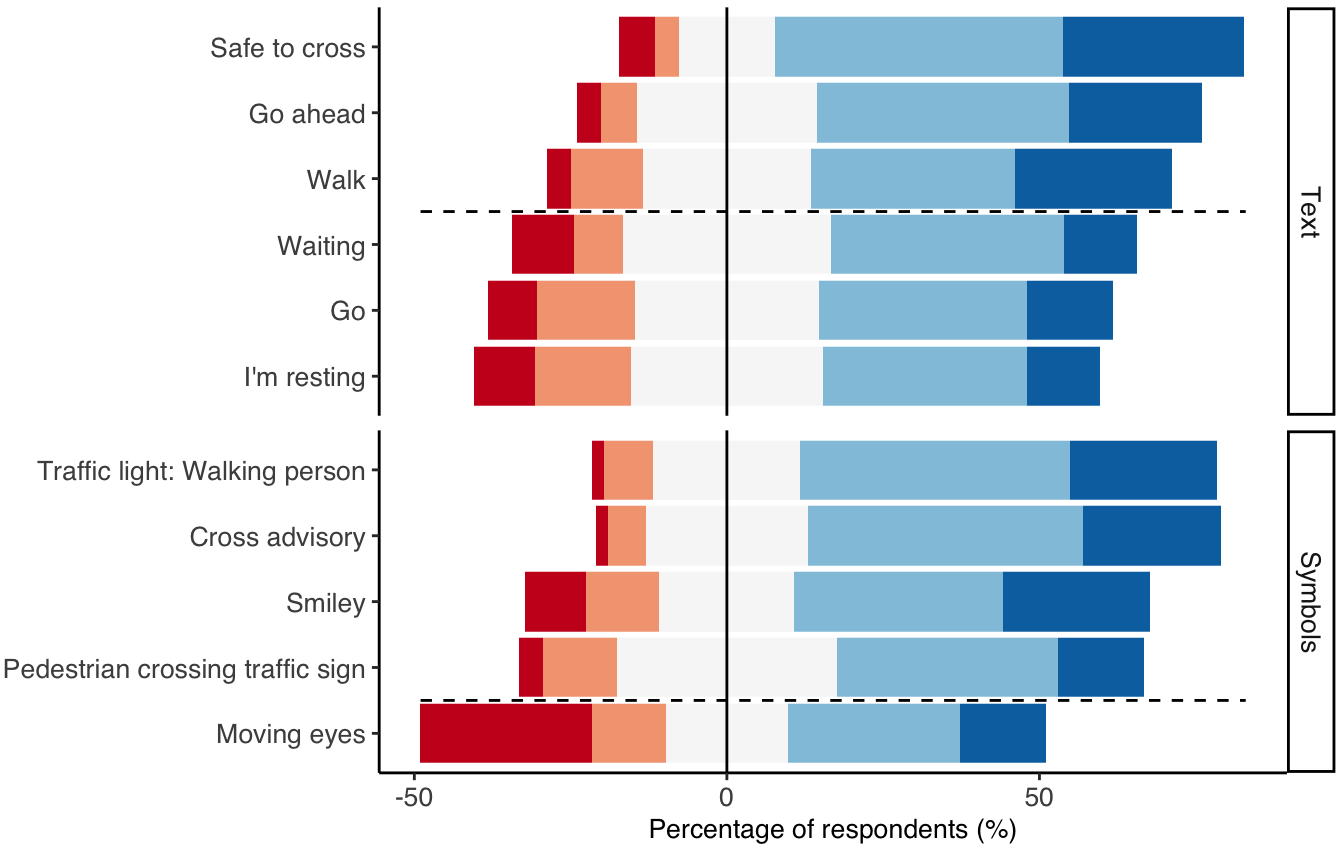}\label{likert_ts_friendly}}
    \end{minipage}\par\medskip
    \centering
    \begin{minipage}{.5\linewidth}
    \centering
    \subfloat[][I believe the \_\_\_ interaction mode communicates clearly]{\includegraphics[width=1.0\textwidth]{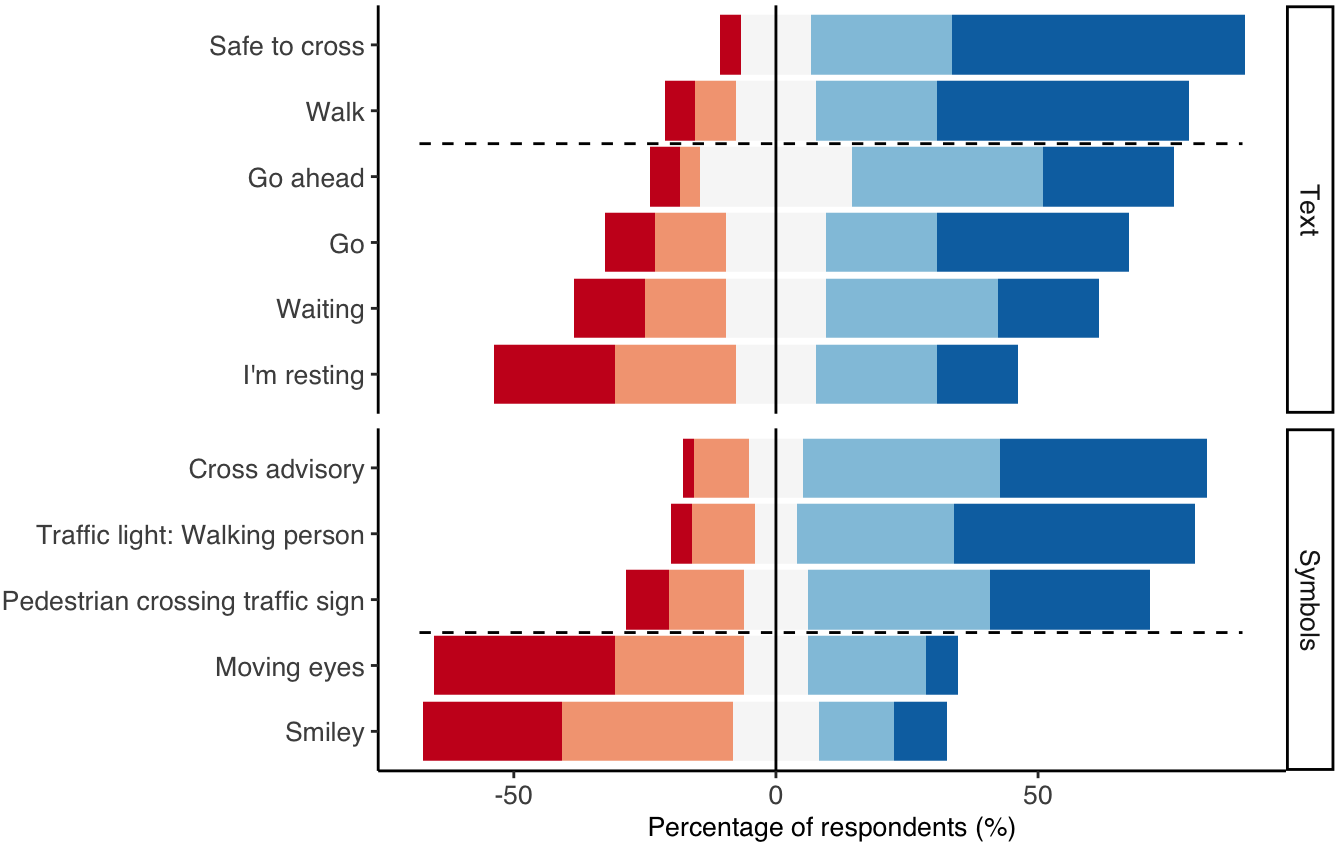}\label{likert_ts_communicates}}
    \end{minipage}%
    \begin{minipage}{.5\linewidth}
    \centering
    \subfloat[][I prefer the \_\_\_ interaction mode over human-driver interaction]{\includegraphics[width=1.0\textwidth]{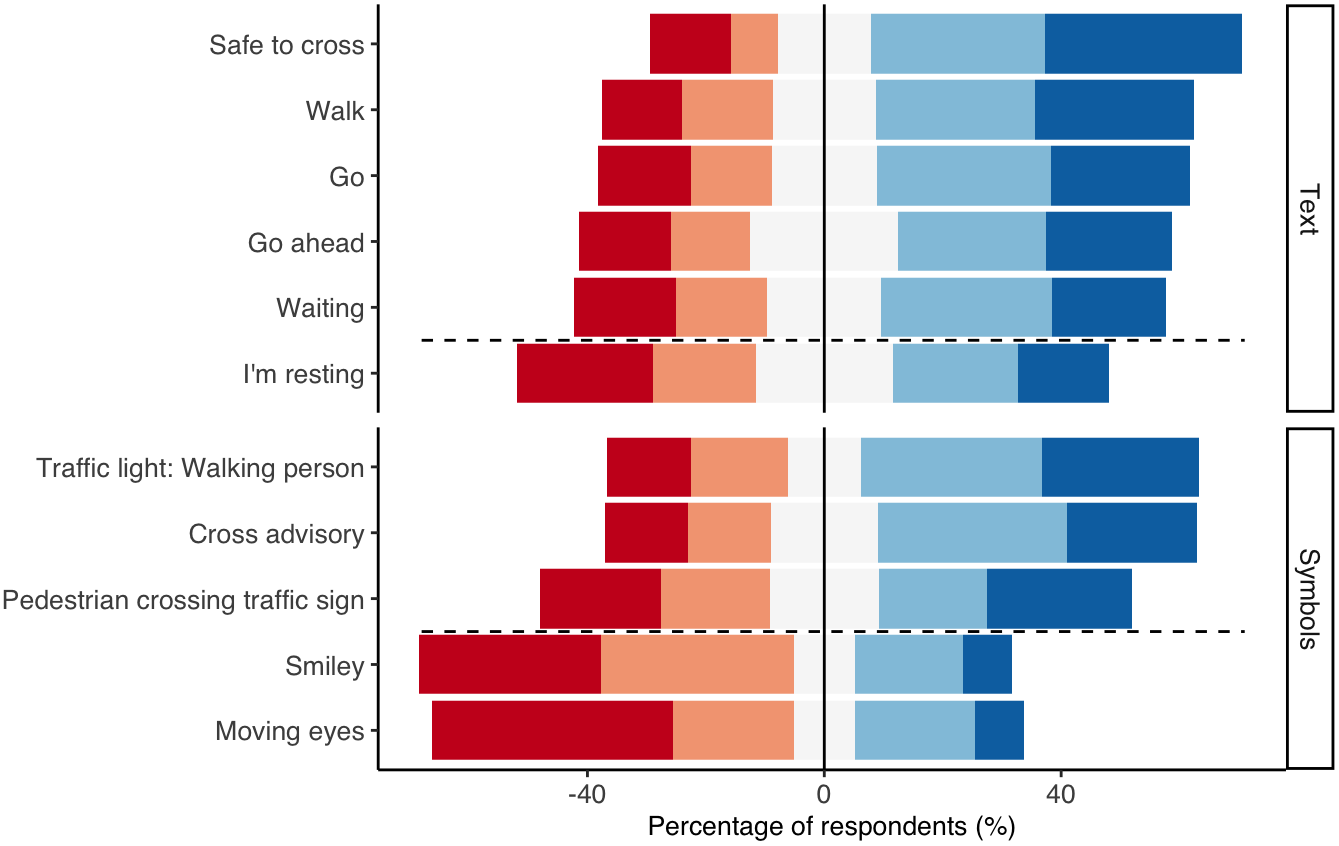}\label{likert_ts_driver}} 
    \end{minipage}\par\medskip
    \vspace{-1em}
    \caption{
    Assessment of the questions \protect\subref{likert_ts_protects}, \protect\subref{likert_ts_friendly}, \protect\subref{likert_ts_communicates} and \protect\subref{likert_ts_driver}
    for the text and symbol interaction mode not driving options. ``Safe to cross" is the highest rated option for the text interaction mode, followed by ``Walk" with no significant difference, and for the symbol interaction mode the walking person as on a traffic light and the cross advisory symbol with no significant difference.
    }
    \label{fig:likerttextsymbol}
  \end{figure*}

Thus far, we have analyzed the rankings of Text, Symbols, Lights and Road Projections regarding our research question.
We now want to look at specific text and symbol options. For this we analyzed the Likert questions, see \mbox{Figure \ref{fig:likerttextsymbol}}.

We will further consider the interaction mode options which are not significant with the highest rated option via the Mann-Whitney U test (Figure \ref{fig:likerttextsymbol}). 
The question ``I believe the \_\_\_ interaction mode protects people from potential risks in the environment" led to the result that the text options ``Safe to cross" and ``Walk", and the symbol options ``Traffic light walking person" and ``Cross advisory" are not significantly different. 
The question ``I believe the \_\_\_ interaction mode looks friendly to the pedestrian" showed that the text options ``Safe to cross", ``Go ahead" and ``Walk", and the symbol options ``Traffic light walking person", ``Cross advisory", ``Smiley" and ``Pedestrian crossing sign" are not significantly different.
Further, the question ``I believe the \_\_\_ interaction mode communicates clearly" showed that the text options ``Safe to cross",  and ``Walk", and the symbol options ``Cross advisory", ``Traffic light walking person" and ``Pedestrian crossing sign" are not significantly different.
The question ``I prefer the \_\_\_ interaction mode over human-driver interaction" showed that the text options ``Safe to cross", ``Walk", ``Go ahead", ``Go" and ``Waiting", and the symbol options ``Traffic light walking person", ``Cross advisory" and ``Pedestrian crossing sign" are not significantly different. 

\vspace{-0.3em}
While the preferred text interaction mode option when the vehicle is not driving for the Likert question is for all four questions ``Safe to cross", the preferred symbol interaction mode option is less clear. Figure \ref{fig:likerttextsymbol} shows that the results for the walking person of a traffic light and the cross advisory symbol are very similar, with only a slight preference for the walking person of a traffic light. 

\subsection{Discussion}
The results of this questionnaire show different important aspects for creating an AV-to pedestrian communication feedback module.
From the questionnaire results we can derive that a combination of text and a symbol when the vehicle is not driving should be used for a feedback module for communicating between a pedestrian and an  autonomous vehicle to increase most legibility, public acceptance and trust, and to identify preference, in the autonomous vehicle’s decision. The text, which increases most legibility, public acceptance and trust, and to identify preference, is ``Safe to cross." However, since the differences between the text interaction modes ``Safe to cross" and ``Walk" are not significant when only one interaction mode is used, the interaction mode ``Walk" could be used as well. The results for the symbol interaction mode option were not as clear, but the symbol options, which increase most legibility, public acceptance and trust, and to identify preference, included a symbol of a walking person.  

\section{Limitations and Future Work} \label{Limitations}
The questionnaire has several limitations. This is because there are too many aspects to consider for a feedback module to ask about each aspect in a questionnaire due to complexity and time constraints. However, this questionnaire was developed as an initial study to reduce the amount of extensive options that could potentially be used as a feedback module.
In this study, we asked about general concepts and therefore omitted, e.g., current law requirements, location, color or size of a possible vehicle-to-pedestrian communication feedback module.
Another limitation is that, although we tried to avoid bias as much as possible by introducing randomization and using a well selected set of possible options, bias in this questionnaire cannot be considered completely excluded. 

We extended the vehicle-to-pedestrian communication feedback module to a vehicle-to-bicyclist communication feedback module, see \cite{schmidt2022comparison}.
In further studies, we will use the results of this questionnaire in a Virtual Reality (VR) user study to create and simulate the selected interaction modes in more detail and in different environments. As a subsequent step we will use the results of the questionnaire and the VR user study to verify in the real-word on a vehicle. 

\section*{Acknowledgments}
This work was supported by the Nevada NASA Space Grant Consortium, Grant No. 80NSSC20M00043.

\bibliographystyle{splncs04}
\bibliography{mybibliography.bib}

\end{document}